     \documentstyle[12pt]{article}%
\begin {document}

     \begin{center}
     \rightline{MRI-PHYS-95/11}
     \vspace{1 cm}

     {\bf Energy  Eigenvalues  For  Supersymmetric   Potentials via Quantum
     Hamilton-Jacobi Formalism}

     \vspace{1 cm}R.S. Bhalla \footnote{$^{}$
     {email: bhalla@mri.ernet.in}},
     A.K. Kapoor \footnote{$^{}$ {email: akksp@uohyd.ernet.in}}
      \vspace{0.5 cm}

     Mehta Research Institute, 10 Kasturba Gandhi Marg, \\Allahabad-211 002,
     India \\


         and \\

     School of Physics, University of Hyderabad,\\ Hyderabad-500 034, India.

     \vspace{0.5 cm}
     and \\
     \vspace{0.5 cm}
     P. K. Panigrahi \footnote{$^{}${email: panisp@uohyd.ernet.in}}
     \vspace{0.5 cm}

School of Physics, University of Hyderabad,\\ Hyderabad-500 034, India.

\vspace{2 cm}

{\bf ABSTRACT}
\vspace{0.5 cm}
\end{center}

     Using quantum Hamilton-Jacobi formalism of Leacock and Padgett, we
     show   how  to  obtain  the  exact  eigenvalues for supersymmetric
     (SUSY) potentials.
\vspace{2 in }

\noindent
July 1995

\newpage
\pagestyle{plain}
\baselineskip = 24pt

\noindent{\bf 1. Introduction}

      In  classical  mechanics, the Hamilton-Jacobi formalism is used to
      arrive at the frequencies of periodic  systems, through the action
      variable,  without  having  to obtain the complete solution of the
      equation   of motion [1]. Analogously,     quantum Hamilton-Jacobi
      (QHJ)  formalism   has   been   developed  [2]   where the quantum
      action variable yields the exact energy spectra of the bound state
      problems,   avoiding  an  explicit solution  of the  corresponding
      Schrodinger  equation.  The  QHJ  equation  is  equivalent  to the
      Schrodinger  equation  and  is  written  in terms of  the  quantum
      momentum  function  (QMF),  $p(x,E)$,  which  is  the  logarithmic
      derivative  of the wave function. The quantization of the energy E
      arises  from  a suitable  contour  integral of the QMF, equated to
      integral multiple of $\hbar$.
      In   this   paper  we  apply  this  formalism  as   developed  by
      Leacock   and   Padgett  to  obtain  exact energy eigenvalues for
      potentials   which    exhibit   supersymmetry  (SUSY)  and  shape
      invariance  (SI)  [3]. It  is  well  known  that  for  potentials
      having    the   above  two  properties,  one  can  find  out  the
      eigenvalues algebraically.

      Our   motivation  in  carrying  out this explicit computation has
      been  (a)  to  gain  a better understanding of the working of the
      QHJ  formalism  and   (b)   to see what kind of singularities are
      possible  for the QMF  which play a  significant role, as will be
      seen later, in determining the eigenvalues. It  is hoped that the
      singularity  structure of QMF will enable us to analyse exactness
      or nonexactness of the well known SUSY WKB formula [3,4].

     The quantum Hamilton-Jacobi equation for a given potential $V(x)$ is
     given by
\begin{eqnarray}  \frac{\hbar}{i}  \frac{  \partial^2  W(x,E)}{  \partial
     x^2}  +  \left(  \frac{\partial W(x,E)}{\partial x}\right)^2& =& E -
     V(x)  \cr& \equiv & p^2_c(x,E),\label{1.2}
\end{eqnarray}
     \noindent
     where  we  have  set  $2m = 1$ and where, $W(x,E)$ is the  quantum
     Hamilton's characteristic  function. The quantum momentum function
     $p(x,E)$ is defined as
\begin{equation}
     p(x,E) \equiv \frac {\partial  W(x,E)} {\partial  x}.
\end{equation}
     \noindent
     The QMF is required to satisfy  the  boundary condition
     $$ \lim_{\hbar \rightarrow 0} p(x,E) \rightarrow  p_c(x,E).$$

\noindent
     Here  $p_c(x,E)$  is  the   classical  momentum  function,  and is
     defined  in  such a way that it has the value  $ + \sqrt{E - V(x)}
     $, just below the branch cut which joins the two turning points.

     Explicitly, the QMF is related to the wavefunction by
     $$p = - i \hbar \,\, \frac{1}{\psi} \frac {\partial \psi}
     {\partial x} \, \, .$$
\noindent
     For  the  $n$-th energy level, $\psi$ has  $n$  zeros  between the
     classical  turning  points.  These  zeros  of  the  wave  function
     correspond  to  simple poles of the momentum function with residue
     given  by  $  \hbar  /i  $ for each of these zeros. The location of
     these  poles is energy dependent and we shall refer to them as the
     moving  poles. Other poles whose location does not depend on energy
     will  be  refered  to  as  the  fixed poles of QMF.  Let  $C$  be a
     contour   enclosing the moving poles between the classical turning
     points. The integral
\begin {equation}
     J(E) \equiv \frac{1} {2\pi} \oint_C p(x,E) dx , \label{1.3}
\end{equation}
 \noindent
     called the  quantum action variable is obviously $J(E)$ is equal
     to $ n \hbar $. Thus
\begin {equation}
     J(E)  = n \hbar,                     \label{1.4}
\end {equation}
\noindent
     when inverted for E, gives the exact energy eigenvalues.

     In this paper we  show  how the QHJ method gives exact eigenvalues
     for  SUSY  potentials.  SUSY   quantum  mechanics  (QM)  has  been
     studied  extensively in  the  last  decade.  For  a review of SUSY
     in  QM, the reader  is referred to recent review article by Cooper,
     Khare  and   Sukhatme  [3].  In  SUSY  QM one  considers a pair of
     potentials defined by ($2m =1$).
\begin{equation}
     V_{\pm} = \omega^2 (x) \pm \hbar \frac{d}{dx} \omega (x),
\end{equation}
\noindent
     where $\omega(x)$ is called the super-potential. The corresponding
     Hamiltonians
     $$ H_{\pm} =   p^2 + V_{\pm}(x),$$
     can be factorized as
\begin{eqnarray}
     H_- &=& A^\dagger A \\H_+ &=& A A^\dagger,
\end{eqnarray}
\noindent
     where
\begin{eqnarray}
      A & =& \hbar \frac{d}{dx} + \omega (x) \\
     A^\dagger & =& -\hbar \frac{d}{dx} + \omega (x).
\end{eqnarray}
\noindent
     Whenever  the  function $\exp \left(-\int^x \omega (y) dy \right)$
     is  square  integrable,  it  represents  the   ground  state  wave
     function of $ H_- $. Let $ \psi_0 $ denote this ground state wave
     function. Then we have
\begin{equation}
     \hspace{.5   in}\omega   (x)   =   -   \hbar   \frac{1}{\psi_0}
     \frac{\partial \psi_0} {\partial x}  \, .
\end {equation}
\noindent
     The  eigenvalues  and  eigenfunctions  for  the two Hamiltonians $
     H_\pm$ are related. Using the intertwining relations $AH_- = H_+A$
     and  $  A^\dagger   H_+   =   H_-  A^\dagger  $,  it can be easily
     shown  that  $E^{(-)}_{n+1}    =   E^{(+)}_{n}$  (apart  from  the
     ground   state   wavefunction   satisfying   $A\psi_0^{(-)} = 0$).
     Hence   $E_{n+1}^{(-)}$    and   $E^{(+)}_{n}$   are   the  energy
     eigenvalues  for  $H_-$ and $H_+$  respectively. It  has  been  known
     that  in  addition  to  SUSY,  if  a  potential   has  a  discrete
     reparameterization   invariance   called   shape  invarience,  the
     corresponding  eigenvalues  and eigenfunctions can  be  explicitly
     obtained  algebraically.  In  the  next section we give  the
     details  of  the  steps in QHJ formalism required  to solve one of
     these potentials. In Sec. 3. we shall take up the solution of a class
     of SUSY potentials. \\

\vspace{1 cm}
\noindent
     {\bf  2. Eckart  Potential} \\
     In  case  of  SI  potentials  both $V_+$ and $V_-$ are of the same
     functional form, albeit with different values of the parameter. In
     the following, we will consider only  the  $V_-$ and apply the QHJ
     formalism   to   obtain   the  corresponding  eigenvalues. In this
     section we shall obtain the energy eigenvalues for bound states of
     the  Eckart  potential given by
     \begin{eqnarray}
     V(x) = A^2 + \frac
     {B^2}{A^2}  +  A(A-\alpha   \hbar) {\rm cosech}^2  \alpha  x - 2 B
     {\rm coth} \, \alpha x   \, . \\
     (x \ge 0) \nonumber
     \end{eqnarray}
 \noindent
     The corresponding superpotential is given by
\begin {equation}
     \omega (x) = - A {\rm coth }\, \alpha x + \frac {B}{A},\label{2.2}
\end{equation}
\noindent
     where  $A$   and  $B$  are  constants  and  the quantum Hamilton-
     Jacobi equation is given by
\begin {equation}
     p^2(x,E) - i\hbar \frac{ \partial p(x,E)}{ \partial  x} = E - A^2
     - \frac {B^2}{A^2} - A(A-\alpha  \hbar) {\rm cosech}^2  \alpha  x
     + 2 B {\rm coth} \, \alpha x \, \, .\label{2.3}
\end{equation}
\noindent
     A simple way to obtain the eigenvalues of Eckart hamiltonian is to
     make use of the transformation
\begin{equation}
     y = \coth \alpha x.
\end{equation}
\noindent
     Under  this  mapping $p(x,E)$ becomes a function of $y$ when $x$  is
     expressed  in  terms  of $y$. We shall continue to use $p$ to denote
     the function of $y$ so obtained. Thus $p(y,E)$ means the QMF $p(x,E)$,
     expressed in terms of the variable $y$. {\it This will be understood
     for all change of variables to be considerd below.} \\

\noindent
     The quantization condition ($\ref{1.4}$) then becomes
\begin{equation}
     I_{C_1}   \equiv  \frac{1}  {2\pi  \alpha}\oint_{C_1}
     \frac{p(y,E) dy}{1-y^2} = n \hbar,\label{123}
\end{equation}
\noindent
     where $C_1$ is the image of the contour $C$ of ($\ref{1.3}$),
     enclosing the turning point,  under the mapping $x \rightarrow y =
     \coth  \alpha x. $ The QHJ equation written in terms of y is given
     as
\begin {equation}
     p^2 - i\hbar \alpha  \frac{dp}{dy} (1+y)(1-y)
     = E - A^2 -\frac {B^2}{A^2} - A(A-\alpha  \hbar) (y +1) (y-1)    +
     2 B y\label{2.5}
\end {equation}
\noindent
     Note  that  the  above  mapping  $y  =\coth  \alpha  r$ introduces
     additional  singularities  in  the  integrand  at  $y \pm 1 $. The
     contour integral $I_{C_1}$ is calculated by deforming the  contour,
     (see  fig  1.)  so  as to enclose all the singular points of the
     integrand. The poles of QMF can be located using available results
     [5]   on  complex  zeroes  of  solutions  of  linear  differential
     equations. They are easily found by inspection for all the problems
     of  interest  in  this  paper.  For  evaluation  of  the  integral
     ($\ref{123}$)   consider  the  contour integral $I_{\Gamma_R}$ for
     a circle $\Gamma_R$  of  radius  $R$ which is taken  to  be  large
     enough so that outside $\Gamma_R$, $p(y)$ has no singularities. The
     singular points of the integrand are the poles at $y = \pm1$ and the
     moving poles enclosed inside $C_1$. Therefore, we have
\begin{equation}
     \frac{1}   {2\pi   \alpha}  \oint_{\Gamma_R}\frac{p  dy}{1-y^2}  =
     \frac{1}  {2\pi \alpha} \left( \oint_{C_1} \frac{p dy}{1-y^2}
     +   \oint_{\gamma_1}   \frac{p   dy}{1-y^2}   +   \oint_{\gamma_2}
     \frac{p dy}{1-y^2} \right).\label{2.4a}
\end {equation}
\noindent
     Denoting   the   integrals   in  ($\ref{2.4a}$) as $ I_{\Gamma_R},
     \, I_{C_1}, \, I_{\gamma_1}$ and  $I_{\gamma_2} $ (fig.1),
     we rewrite ($ \ref{2.4a} $) as
\begin{equation}
     I_{\Gamma_R}=I_{C_1}    +    I_{\gamma_1}   +   I_{\gamma_2}.
     \label{15z}
\end{equation}
\noindent
     The   contour  integral  $I_{\Gamma_R}$  is calculated by one more
     change of variable to $z = 1/y$. In terms of the variable  $z$ the
     integral  $I_{\Gamma_R}$  becomes
     \begin{eqnarray}
     I_{\Gamma_R} &=&   \oint_{\Gamma_R} \frac{p dy}{1-y^2} \\
     & = & \oint_{\gamma_0}\frac{p dz}{1-z^2}  \\
     & \equiv & I_{\gamma_0} \label{17z}
     \end{eqnarray}
\noindent
     where   $\gamma_0$   is  a small circle in the $z$-plane enclosing
     only one singular point  $z  = 0$. It is worth reminding that both
     the contours  are  in  the  anticlockwise  direction  and  the
     singularity at $ y \rightarrow \infty $ is mapped to the singularity
     at $ z = 0$.

 \noindent
     Therefore ($\ref{123}$), ($\ref{15z}$) and ($\ref{17z}$) give
\begin{equation}
      n \alpha \hbar = \frac{1}{2 \pi}\oint_{ C_1}
     \frac {p dy} {1-y^2} = \frac {1}{2\pi}\left(
     \oint_{\gamma_0} \frac{p dz}{1-z^2} - \oint_{\gamma_1}
     \frac{p dy}{1-y^2}  -  \oint_{\gamma_2}  \frac{p dy}{1-y^2} \right).
     \label{2.5a}
\end{equation}
\noindent
     The  calculation  of  various  integrals  requires behavior of the
     momentum function near the singular points. In particular, we need
     the  value of the function at $y = \pm 1$ and its residue at $ z =
     0$.  These  are  calculated  by substituting appropriate Taylor or
     Laurent  series  expansion of $ p$ in the QHJ equation and solving
     for  the  first  few  coefficients  in  the  series expansion. For
     example,  for  calculation  of  the contour integral around $ y  =
     \pm   1$, the  series expansion of $p(y,E)$ around $ y  =  \pm  1$
     is  used.  The  integrand  suggests that we need to calculate only
     the  coefficient  of  the  constant  term.  We illustrate this for
     computing  the  integrals  $I_{\gamma_1}$;  for this purpose we at
     first expand $p(y,E)$ as
     $$
     p(y,E) = \alpha_0 + \alpha_1 (y - 1) + \alpha_2 (y - 1)^2 + \cdots
     \cdots
     $$
\noindent
     and  substite  the expansion in ($\ref{2.5}$). Comparing the terms
     independent of $y$ on both sides gives
\begin{equation}
     \alpha^2_0 = E - A^2 - \frac{B^2}{A^2} + 2B \label {2.7}
\end{equation}
\noindent
     Let $\beta_0$ be the constant term in the expansion of $p(y,E)$ in
     powers  of  $(y+1)$. Then $\beta_0$ is similarly determined and is
     given by
\begin{equation}
     \beta^2_0 = E - A^2 - \frac{B^2}{A^2} - 2B \label {2.8}
\end{equation}
\noindent
     Note  that,  there  are  two  roots  for $\alpha_0$ and $\beta_0$,
     corresponding  to  the  two  signs of the square of the right hand
     side   in   $(\ref   {2.7})$and $(\ref {2.8}) $.
     Similarly for obtaining the integral $I_{\gamma_0}$, we need to
     compute the residue of $p(z,E)$ at $z = 0$.  We expand $p(z,E)$ as
\begin{equation}
      p(z,E) =  b_1/z + a_0 + a_1 z + \cdots \label{2.9z}
\end{equation}
\noindent
     and substitute ($\ref{2.9z}$) the QHJ equation written in terms of
     the variable $z$
\begin{equation}
       p^2- i\alpha \hbar\left( \frac{1-z^2} {z^2}\right) \frac{dp}{dz}
       =E   -   A^2   -   \frac    {B^2}{A^2}   -   A(A-\alpha   \hbar)
       \left(\frac{1 -z^2}{z^2}\right) + \frac{2B} {z} .
       \label {2.6}
\end{equation}
     Comparing the coefficients of $1/z^2$ on the two sides gives,
\begin{equation}
     b_1 = \frac{-i \alpha \hbar \pm i(\alpha \hbar-2A)}{2}
     \label{132}
\end{equation}
\noindent
     We select the correct root for $\alpha_0, \beta_0$ and $b_1$ from
     ($\ref{132}$), ($\ref{2.7}$) and ($\ref{2.8}$) by comparing them with
     the answer for $E=0$. It is straight forward to obtain the value of
     $b_1$, $\alpha_0$ and $\beta_0$  for  $E = 0$, by recalling that for
     zero energy the  QMF is related to the superpotential by $p(x,E=0) =
     i \omega (x)$. Writing the superpotential $\omega (x) $ in terms of
     $z$, we get
\begin{equation}
     \omega (z) =  -\frac{A}{z} + \frac{B}{A}.
\end{equation}
\noindent
      The  residue of $\omega(z)$ at $z = 0$ is $ - A $. Comparing this
      answer  with  value  of  $b_1$, we see that the correct choice of
      $b_1$ is given by $ b_1 = - i A $  for all $E$. Similarly one looks
      for the coefficient of expansion of $w$ in powers of $y\pm1$ and
      this is compared with the values of $\alpha_0$ and  $\beta_0$
      respectively for $E=0$. It is found that these roots have relative
      opposite signs and are given by
      \begin{eqnarray}
      \alpha_0 &=& -\sqrt{E - A^2 - \frac{B^2}{A^2} + 2B} \label {2.7z} \\
     \beta_0 &=& \sqrt{E - A^2 - \frac{B^2}{A^2} - 2B} \label {2.8z}
      \end{eqnarray}
\noindent
     It  may  be noted that the correct sign of the residues calculated
     can also  be  fixed by taking $ \hbar \rightarrow 0$ and looking at
     the behaviour of $p_c$ near the point of interest.  This procedure,
     as originally suggested by Leacock and Padgett, is a bit complicated.
     For SUSY potentials under consideration in this article we have found
     it useful to follow the alternate  procedure as explained above.

     The   contour   integrals   $I_{\gamma_0}$,   $I_{\gamma_1}$   and
     $I_{\gamma_2}$ are, therefore, computed to be
     $$
     I_{\gamma_1}  =  \frac{i  \alpha_0}{2 \alpha}, \, \hspace{.5 in} \,
     I_{\gamma_2}  =  \frac{i  \beta_0}{2  \alpha}, \, \hspace{.25 in}
     {\rm and} \hspace{.25in} \, I_{\gamma_0} = \frac{A}{\alpha}. $$
     \noindent
     Thus the energy eigenvalues are obtained from
\begin{equation}
     I_{C_1} = n \hbar = \frac{A}{\alpha} -\frac{i \beta_0}{2 \alpha}
     - \frac{i \alpha_0}{2 \alpha}
\end{equation}
     \noindent
     which on further simplification gives
\begin{equation}
     E_n = A^2 + \frac{B^2} {A^2} - \frac{B^2}{(n\alpha \hbar + A)^2 }
            -(n\alpha \hbar + A)^2 \, \, \, ,
\end{equation}
\noindent
     One  can  arrive  at  the  above  energy  eigenvalues using  other
     mappings.  Use  of  a  different mapping will be considered in the
     next  section and we shall then show how to obtain the bound state
     eigenvalues for other SUSY potentials.\\
\newpage
\noindent
{\bf 3. Other SUSY potentials } \\
      In   this  section  we  will show  how the QHJ method could be used
      for  other   SUSY   potentials. We will use $y = \exp (i \alpha x)$
      mapping for the SUSY potentials involving trigonometric  functions.
      The remaining potentials involving hyperbolic functions $y =\exp
      (\alpha x) $  will be used.  The treatment of each of these cases
      runs  parallel  to the treatment given to the Eckart  potential  in
      the previous section except for a new  point  which requires special
      attention. The use of the mapping $y = \exp (\alpha x)$ gives  rise
      to extra energy dependent poles in  $p(x,E)$ in  the non-classical
      region. Computation  of the exact energy eigen-values  requires the
      knowledge of an integral along a contour enclosing   these  energy
      dependent poles. In  all  the  cases investigated in this paper this
      integral can  be related to the integral  around   the contour which
      encloses  poles  in the  classical region on the real axis.

      In  this  section  at first we shall work out the eigenvalues for
      the Eckart potential again, using the mapping $ y  = \exp (\alpha
      x)$, but concentrating  only  on  the  new  points as compared to
      the treatment in  the previous section.

\noindent
     The corresponding super potential for the Eckart potential written
     in terms of the variable $y = \exp ( \alpha  x) $ is
\begin{equation}
     \omega(y) = -A \left( \frac{y^2 +1}{y^2-1} \right)+ \frac{B}{A},
\end{equation}
\noindent
      and the QHJ equation is
\begin{equation}
     p^2 - i\hbar \alpha y  \frac{dp}{dy}  = E - A^2 -\frac{B^2}{A^2}
     - \frac{4A(A-\alpha  \hbar) y^2}{(y^2-1)^2}
     + \frac{ 2 B (y^2 +1)} {(y^2 -1)}.\label{4.1a}
\end{equation}
\noindent
 Now the equation
\begin{equation}
      E - A^2 -\frac{B^2}{A^2} - \frac{4A(A-\alpha  \hbar) y^2}{(y^2-1)^2}
        + \frac{ 2 B (y^2 +1)} {(y^2 -1)} =0 ,\label{4.1}
\end{equation}
     has four solutions for the turning points. These are shown as $A$,
     $B$,  $A^\prime$  $B^\prime$ (see Fig. 2). The moving poles of $p$
     are  (i)  on the real axis and between $A$ and $B$ and (ii) on the
     real axis between $A^\prime$ and $B^\prime$.  It should however be
     noted, that  two of these turning points $A^\prime$ and $B^\prime$
     are  in  the  non-classical   region. We note that the symmetry $y
     \rightarrow - y$ in ($\ref{4.1}$) interchanges $A$ with $A^\prime$
     and   $B$   with  $B^\prime$. This  symmetry  implies
\begin{equation}
     I_{C_1} = I_{C_2} ,
     \label{139}
\end{equation}
     where $C_1$ and $C_2$ are contours enclosing $A, B,$ and $ A^\prime,
     B^\prime$ respectively as shown in Fig. 2. Next note  that now $y = 0$
     is a pole of the integrand in the action integral (see ($\ref{239}$)
     below). Also $p(y)$ has  poles at $y = 1 $ and $y  = -1$ because the
     right hand side of QHJ  is singular at $ y= 1$ and $ y=-1.$ Introducing
     $I_{\Gamma_R}$ for the large circle $\Gamma_R$ enclosing all the
     singular points we arrive at
\begin{equation}
     I_{\Gamma_R} = I_{C_1} + I_{C_2} + I{\gamma_1} + I_{\gamma_2} +
                            I_{\gamma_3} .
\label{234}
\end{equation}
Here $I_{\gamma_1}, I_{\gamma_2},I_{\gamma_3}$ are integrals along
contours $\gamma_1, \gamma_2, \gamma_3$ which enclose the singular points
$y=1, y=-1$ and $y=0$ respectively. Therefore, using ($\ref{139}$), the
quantization  condition
\begin{equation}
     I_{C_1} \equiv \frac{1}{2\pi \alpha} \oint_{C_1}\frac{dy}{y} p(y,E)
     = n \hbar.
     \label{239}
\end{equation}
\noindent
    becomes
\begin{equation}
     I_{\Gamma_R} - \sum_p I_{\gamma_p} = 2n \hbar.
     \label{567}
\end{equation}
\noindent
The rest of the calculation is same as in Sec. 2. and one easily arrives at
\begin{equation}
     E_n = A^2 + \frac{B^2} {A^2} - \frac{B^2}{(n\alpha \hbar  + A)^2 }
                       -(n\alpha \hbar + A)^2 .
\end{equation}
\noindent
      The calculation of eigenvalues for other SUSY potentials proceeds
      in  a similar fashion. The results are summarized in Table I and
      Table II. The range of the variable $x$ is $ - \infty < x < \infty $,
      unless indicated otherwise. Expressions listed in the coloumn four
      are $\alpha I_{\gamma_p}$ are the values for a contour  $\gamma_p$
      enclosing only the singular point indicated in the third column of
      the table. The value of $I_{\gamma_p}$  for the  pole  at $\infty$
      stands for the value $I_{\Gamma_R}$. For the fixed poles at $y = 0$
      and $\infty$ the square roots in the residue are found to have
      relatively opposite signs, on comparision with the coefficients of
      corresponding terms in the expansion of $\omega (x)$ for $E=0$. The
      eigen-values listed in the last column are obtained using
      ($\ref{567}$).

\vspace{1 cm}
\noindent
     {\bf 4. Conclusion: }  \\
     In  conclusion, we have explicitly worked out the eigenvalues  for
     the    SUSY  potentials  using the QHJ method. Apart from checking
     the   correctness  of   the  formalism,   this  exercise  provides
     insight    into   solvability   of    these  potentials.  The main
     effort  involved  in  use  of  this  scheme  lies in selecting the
     correct   root  for  the residues needed. This problem was  solved
     here by comparing the answers obtained from the QHJ for $E=0$ with
     that  obtained  from  the superpotential. In general the choice of
     the  correct  root  for  the  residue  can  be   made by using the
     boundary  condition on QMF, viz., $p \rightarrow p_c$ in the limit
     $\hbar \rightarrow 0$ and where the branch of $p_c$ is selected in
     such  a   way that it corresponds to the positive sign just  below
     the cut on the real axis joining  the   physical turning  points.

     The  Leacock-Padgett method is a powerful method for obtaining the
     eigenvalues analytically as well as numerically and can be applied
     to  other  potentials. It is  worth  pointing  out  that  for  the
     potentials considered  in this article, the SUSY-WKB approximation
     gives  exact answer. Since  QHJ  as  shown  here  also gives exact
     answer,   it  is natural to enquire  as  to  the  relation between
     these   two  approaches.This  will  throw light on the question of
     exactness    of  the  SUSY WKB formula. This investigation will be
     reported elsewhere.

     {\it  Acknowledgements  }  :  We  acknowledge  useful  discussions
     with   Profs.  A. Khare, V. Srinivasan, Pankaj Sharan, U. Sukhatme
     and  S.  Chaturvedi and Drs. C.  Nagraj Kumar and R. Adhikari. Two
     of  us  (R.S.B.   and   A.K.K.)  would  like  to  thank  Director,
     Mehta   Research   Institute   for  hospitality and support during
     their stay.

\vspace{1 cm}
\noindent{\bf References}
\begin {enumerate}
     {\item  H. Goldstein, Classical Mechanics (Addison-Wesley, New
     York,1950).}

     {\item  R.A.  Leacock and M.J. Padgett, Phys. Rev. {\bf D28},
      2491 (1983)\\R.A.Leacock and M.J. Padgett, Phys. Rev. Lett.
     {\bf 50}, 3 (1983).}

     {\item  F. Cooper, A. Khare and U. Sukhatme Phys. Reports
     {\bf 251}, 267 (1995).}

     {\item A. Comtet, A. Bandrauk and D.K. Campbell, Phys. Lett.
     {\bf B150}, 159 (1985). }

     {\item E. L. Ince , {\it Ordinary Differential Equations}, Dover
Publcations N.Y
     (1956). }
\end {enumerate}

\pagebreak

\pagestyle{empty}
\topmargin=-1truein
\textwidth=6truein
\textheight=9truein
\flushleft
\setlength{\oddsidemargin}{-.5 in}
\setlength{\evensidemargin}{.5 in}

Table 1 :  Trignometric Potentials. \\
For all the potentials listed in this table the mapping  $y =\exp( i \alpha x
)$ is used.\\
\vspace{.5cm}
\begin{tabular}{c c c c c }
\hline \hline
Name of & Potential & Location of
& $\alpha I_{\gamma_p}$ & Eigen \\
potential & &  fixed poles
&  & value \\
\hline
 & & & & \\
  &  &   0
& $   \sqrt{E+A^2}$ &   \\

& & & & \\
Scarf I  &
 $ -A^2 + (A^2+B^2+A\alpha \hbar)\times$ &
  $i$   & $ -A+B $ & $ (A + n\alpha \hbar )^2 $    \\
(Trignometric)& $\sec^2\,\alpha x - B(2A + \alpha \hbar) \times $ &
 & & $- A^2 $ \\
 & $ \sec\,\alpha x \, \tan\,\alpha x$ & $-i$
& $ -(A + B) $ &
\\ & ($- \pi/2 \le \alpha x \le \pi /2 $)  & & & \\
 &  &  $ \infty $   & $ - \sqrt{E+A^2}$ &   \\
 & & & & \\
 & & & & \\
 & & & & \\
 &  &  0
& $ - \sqrt{E+A^2 -B^2/A^2 +2iB}$ &    \\
  & & & & \\
Rosen- & $ A(A -\alpha \hbar ) {\rm cosec}^2\,\alpha x $
  &  $1$   & $  A $ &  $+ B^2/A^2$ \\
Morse-I & $- A^2 + B^2/A^2  $ & & &  $ A^2 -(A +  n\alpha \hbar )^2  $ \\
 (trignometric) & $ + 2B \cot \,\alpha x$
 &  $-1$ & $ A $ & $ -B^2/(A+n\alpha \hbar)^2$
\\   & ($0 \le \alpha x \le \pi$) & & & \\
 &  &  $\infty $ & $  \sqrt{E+A^2 -B^2/A^2 -2iB}
$ &   \\
 & & & & \\
\hline \hline
\setlength{\oddsidemargin}{0. in}
\setlength{\evensidemargin}{0. in}
\end{tabular}
\pagebreak

Table 2 :  Hyperbolic Potentials. \\
For all the potentials listed in this table the mapping  $y=\exp(\alpha x)$
is used. \\
\samepage
\begin{tabular}{c c c c c c }
\hline \hline
Name of & Potential & Location of
& $\alpha I_{\gamma_p}$ & Eigen \\
potential & &  fixed poles
&  & value \\
\hline
 & & & & \\
 &  &   0 & $ -i \sqrt{E-A^2}$ &    \\
&  & & & \\
Scarf II  & $A^2 + (B^2-A^2-A\alpha \hbar)\times $ &  $i$ &
$ iB -A$ &   \\
(hyperbolic) &
$ {\rm sech ^2}\,\alpha x + B(2A + \alpha \hbar ) \times $
 & & & $ A^2 -(A -  n\alpha \hbar  )^2 $   \\
 & $ {\rm sech}\,\alpha x \, \, {\rm tanh} \,\alpha x$
   & $-i$   & $ -iB -A$ &   \\
 &  & & & \\
 &  &   $ \infty $   & $ i \sqrt{E-A^2} $ &   \\
 & & &  & \\
 & & &  & \\
 &  &   0 & $ - i \sqrt{E-A^2 -B^2/A^2 +2B}$ &  \\
 & &  & &  \\
Rosen - & $A^2 + B^2/A^2$  &   $i$   &  $ - A $ &
$ A^2 + B^2/A^2  $ \\
  Morse II & $ -A(A +\alpha \hbar ) \, {\rm sech^2} \,\alpha x $
& & & $ - (A -  n\alpha \hbar )^2 $ \\
(Hyperbolic) & $ + 2B {\rm tanh} \,\alpha x$
 &  $-i$   & $ - A $ & $ - B^2/(A-n\alpha  \hbar)^2$   \\
& & & &  \\
 &  &  $\infty $ & $ i \sqrt{E-A^2 -B^2/A^2 - 2B}
$ &   \\
 & & & & \\
 & & & & \\
&  &  0 & $  i \sqrt{E-A^2 -B^2/A^2 - 2B} $ &   \\
& & & & \\
Eckart & $A^2 + B^2/A^2$
 &  $1$   & $  A $ & $ A^2 + B^2/A^2$   \\
(Hyperbolic)  & $ +A(A -\alpha \hbar) \, {\rm cosech}^2\,\alpha x $ &
 & & $ -(A +  n\alpha \hbar )^2  $  \\
 & $ - 2B {\rm coth}\,\alpha x$
 &  $-1$ & $ A $ & $ -B^2 / (A+n\alpha \hbar)^2$
\\   & $(x \geq 0 )$& & &  \\
 &   & $\infty $ &$ - i  \sqrt{E-A^2 -B^2/A^2 +2B}$   &   \\
 & & & & \\
 & & & & \\
 &  &   0
& $ - i \sqrt{E-A^2}$ &   \\
& & & & \\
Generalised&
$A^2 + (B^2+A^2+A\alpha \hbar )\times$
 &  $1$   & $ -A + B $ &   \\
Poschl-& ${\rm cosech}^2\,\alpha x - B(2A + \alpha \hbar ) \times $&
 & & $ A^2 -(A -  n\alpha \hbar )^2  $ \\
Teller & $ {\rm coth} \,\alpha x \, {\rm cosech}\,\alpha x $
   & $-1$   & $ -(A+B) $ &
\\ &$( x\geq 0)$ & & \\
 &  &  $ \infty $   & $
i \sqrt{E-A^2} $ &   \\
 & & & & \\
 \hline
 \hline
\end{tabular}

\pagebreak

\end{document}